\documentclass[preprint]{aastex}

\begin{document}

\title{Star Formation Properties in Barred Galaxies(SFB). I. UV-to-Infrared Imaging \& Spectroscopic Studies on NGC~7479}
\author{Zhi-Min Zhou\altaffilmark{1,2,3}, Chen Cao\altaffilmark{4,5}, Xian-Min Meng\altaffilmark{1}, and Hong Wu\altaffilmark{1,3}}
\altaffiltext{1}{National Astronomical Observatories, Chinese Academy of Sciences, Beijing 100012, China; zmzhou@bao.ac.cn, mengxm@bao.ac.cn, hwu@bao.ac.cn. }
\altaffiltext{2}{Graduate School, Chinese Academy of Sciences, Beijing 100039, China.}
\altaffiltext{3}{Key Laboratory of Optical Astronomy, National Astronomical Observatories, Chinese Academy of Sciences, Beijing 100012, China.}
\altaffiltext{4}{School of Space Science and Physics, Shandong University at Weihai, Weihai, Shandong 264209, China; caochen@sdu.edu.cn.}
\altaffiltext{5}{Shandong Provincial Key Laboratory of Optical Astronomy \& Solar-Terrestrial Environment, Weihai, Shandong, 264209, China.}
\shortauthors{Zhou, Cao, Meng \& Wu}
\shorttitle{Multi-wavelength Study of NGC~7479}

%% Abstract
\begin{abstract}

Large-scale bars and minor mergers are important drivers for the secular evolution of galaxies. Based on
ground-based optical images and spectra as well as ultraviolet data from the {\it Galaxy Evolution Explorer} and infrared data from the {\it Spitzer Space Telescope}, we present a multi-wavelength study of star formation properties in the barred galaxy NGC 7479, which also has obvious features of a minor merger. Using various tracers of star formation, we find that under the effects of both a stellar bar and a minor merger, star formation activity mainly takes place along the galactic bar and arms, while the star formation rate changes from the bar to the disk. With the help of spectral synthesis, we find that strong star formation took place in the bar region about 100 Myr ago, and the stellar bar might have been $\sim$10 Gyr old. By comparing our results with the secular evolutionary scenario from Jogee et al., we suggest that NGC 7479 is possibly in a transitional stage of secular evolution at present, and it may eventually become an earlier type galaxy or a luminous infrared galaxy. We also note that the probable minor merger event happened recently in NGC 7479, and we find two candidates for minor merger remnants.
\end{abstract}


%% Sec. 1
\section{Introduction}
The physical processes of galaxy evolution can be classified into fast and slow according to their timescales, and also can be divided into internal and external based on origins of drivers \citep{KK04}. As the Universe expands, internal secular processes will become important. Secular evolution is different from dissipative collapses and mergers of galaxies, which are rapid and violent \citep[e.g.,][]{Sandage90,Sandage05}. Secular processes can make the energy and mass rearrange, and be driven by galactic structures such as bars, oval disks, spiral arms, triaxial dark halos inside of galaxies \citep{Athanassoula02, KK04, Kormendy05, Kormendy08}, also by minor mergers and other environmental factors outside the galaxies \citep{Bournaud05, Bournaud07, Jogee06}.

Many observations and researches have shown that at low redshift, especially in the local universe (z $\sim$ 0), mergers are less common \citep[e.g.,][]{Le00, ConseliceB03}, so secular evolution will have important effects on galaxies \citep{KK04}. Pseudobulges, galactic structures produced by secular evolution, are common in disk galaxies \citep{Kormendy05, Kormendy08}, and also indicates that this evolution is a common phenomenon to galaxies. This process is not only limited to low redshift galaxies, e.g., \citet{Genzel08} found that secular evolution may help account for the stellar buildup observed in massive galaxies at z $\sim$ 2. In addition, as the Universe expands, secular evolution will take the place of mergers to dominate the evolution of galaxies \citep{KK04}.

As important internal structures to secular evolution, galactic stellar bars have been discussed in many works, both theoretical and numerical \citep[e.g.][]{Sellwood99, Barazza07, Athanassoula09}. The bar is a common phenomenon in the universe because about 65$\%$ of nearby spiral galaxies have bars, among which over 30$\%$ have strong bars \citep{Eskridge00}, and the percentage of bars remains high out to redshift z $\sim$ 1 \citep{Elmegreen04, Jogee04}. Bars can provide intense nonaxisymmetry in the gravitational potential, which will make gas lose angular momentum and then fall down to the inner region of galaxies \citep{Athanassoula03, Kormendy05}. \citet{Sheth05} compared the molecular gas distribution in a sample of nearby galaxies from the BIMA CO (J=1-0) Survey of Nearby Galaxies (SONG), and found that more molecular gas is concentrated in the central kpc of barred spirals than that in other Hubble type galaxies, consistent with radial inflow driven by the bar.
After the gas density becomes supercritical in the central parts, the next step is to trigger circumnuclear star formation \citep{Sheth05}, and (pseudo)bulges would also grow in this progress \citep{KK04}. These have been proved by a large number of observations \citep[e.g.,][]{Knapen06, Shi06, Gadotti10} and simulations \citep[e.g.,][] {Athanassoula09}.
\citet{Gadotti01} also studied the broadband UBV color profiles for 257 Sbc barred and nonbarred galaxies, and found the bulges of barred galaxies are bluer than others, indicating an increase of the star formation rate (SFR) in the central regions of these objects. \citet{Regan06} found four barred galaxies of 11 spiral galaxies have strong central excesses in both 8 $\mu$m and CO emission. Whereas \citet{Ho97a} found that the same effect only in early-type barred spirals using the luminosity and the equivalent width of H$\alpha$ emission. \citet{Fisher06} found that some barred galaxies don't have higher SFRs than those of other types in a sample of 50 galaxies spanning Hubble types E to Sc using Spitzer infrared color profiles.

Similar to stellar bars, minor mergers also can drive gas into the galactic central region and fuel nuclear activities \citep{Jogee06}. Observations and N-body simulations have shown that minor mergers can lead to high SFR \citep{Ferreiro04, Ferreiro08, Kaviraj09}, engender circumnuclear rings \citep{Knapen04, Mazzuca06}. In addition, the evolution of stellar bars may relate to minor mergers in numerical simulations \citep{Berentzen03, Emilio08}. However, these correlations still need to be studied further.

In order to better understand the effects of bars on the secular evolution of galaxies, it is essential to study the star formation properties of nearby barred galaxies. Thus we selected a barred galaxy sample from three large Spitzer Legacy or GTO programs: the Spitzer Infrared Nearby Galaxies Survey \citep[SINGS;][]{Kennicutt03}, the Local Volume Legacy Survey \citep[LVLS;][]{Lee08}, and the Infrared Hubble Atlas \citep{Pahre04}. It is consisted of 73 nearby (z $<$ 0.02) barred galaxies (including Hubble types SB \& SAB). One most interesting object in this sample is NGC~7479. It is a SBbc galaxy \citep{de91} with a large stellar bar and two strongly asymmetric spiral arms, among which the western one is much stronger. Its small and bright nucleus is classified as LINER \citep{Keel83} and Seyfert 1.9 \citep{Ho97b}. Observations and simulations have been made in the studies of NGC~7479 \citep[e.g.,][]{Martin97, LaineH99, Aguerri00, Huttemeister00, Laine06, Laine08}. There are amount of molecular gas along the stellar bar and in the central region of NGC~7479, and also obvious dust lanes and large-scale shocks along the bar \citep{Laine98}. \citet{Rozas99} showed that intense star formation regions can also be found in the barred region. \citet{Huttemeister00} found that there may be a nuclear ring, a torus, disk or inner bar in the bulge of this galaxy. It seems that this galaxy has no close companions although it looks like an interacting system from its morphology \citep{Laine98, Saraiva03}, so it may have suffered a minor merger recently \citep{Laine08}. Based on the facts above, the secular evolution driven by bar and minor merger simultaneously make NGC~7479 an interesting galaxy worth careful study.

In the present paper, we study the properties of NGC~7479, mainly focusing on the bar and star formation regions based on multi-wavelength data from ultraviolet (UV) to infrared (IR). It is organized as follows: Section 2 describes the data we used and their reductions. Section 3 presents the main results from this study. Section 4 gives some discussions, and we made a summary in section 5.

%% Sec. 2
\section{Data Acquisition and Reduction}
\label{data}
We archived FUV and NUV images \citep{Gil07} from the NASA Extragalactic Database (NED\footnote{http://nedwww.ipac.caltech.edu/}) and infrared images (P.I. Giovanni Fazio) from the Spitzer Space Telescope data archive using the Leopard software. In addition, we observed NGC~7479 using ground-based telescope and obtained its optical images and spectra. The detailed information has been listed in Table~\ref{tabl1}.
%% Sec. 2.1
\subsection{UV Data}
We obtained the UV images of NGC~7479 observed using Galaxy Evolution Explorer \citep[GALEX;][]{Martin05}. The galaxy was imaged in both far-ultraviolet (FUV; 1350$-$1750\AA) and near-ultraviolet (NUV; 1750$-$2750\AA) broadband. These images are from the Nearby Galaxies Survey \citep[NGS;][]{Gil07}, which includes observations of nearby galaxies of different types and environments based on GALEX, and they are reduced through the GALEX pipeline. We made constant sky backgrounds subtracted and flux calibration using keywords of mean sky-background level (SKY) and zero point in AB magnitude scale (ZP) in the headers of FUV and NUV images. The final images are characterized by a point-spread function (PSF) with a FWHM of $\sim$6$''$ and a pixel size of 1.5$''$.

%% Sec. 2.2
\subsection{Optical Data}
We observed NGC~7479 using the 2.16m telescope at Xinglong Observatory of the National Astronomical Observatories of the Chinese Academy of Sciences\footnote{http://www.xinglong-naoc.org/English/216.html}. On September 12, 2009 the observation was carried out by the BAO Faint Object Spectrograph and Camera (BFOSC) with a Lick 2048$\times$2048 CCD detector, whose field of view is about 10$'$$\times$ 10$'$, along with a pixel size of 0.305$''$. There were four filters used, of which the most important one was the narrowband interference filter centered near the redshifted H$\alpha$ emission line with central wavelength at $\sim$6610\AA~and FWHM of 70\AA~(marked as H$\alpha$2). Besides, three other filters were broadband B, V, and R, respectively. The exposure time was 1200s in B, 1100s in V, 600s in R and 3000s in H$\alpha$2, respectively.

The reduction of these optical images was the standard CCD reduction pipeline, including checking images, adding keywords to fits headers, subtracting overscan and bias, correcting bad pixels, flat-fielding and removing cosmic-rays. All of these were applied to images by IRAF\footnote{IRAF is the Image Reduction and Analysis Facility written and supported by the IRAF programming group at the National Optical Astronomy Observatories (NOAO) in Tucson, Arizona which is operated by AURA, Inc. under cooperative agreement with the National Science Foundation}. The absolute flux calibration of broad-band images were made using the standard star GD246, which was selected from the Landolt fields \citep{Landolt92} and observed with corresponding filters at the same night. The final flux conversion factors are 4.90, 1.16, 1.25 10$^{-21} erg\ s^{-1} cm^{-2} \AA^{-1}$/DN (DN is Digital Numbers) for B, V, R bands, respectively. The uncertainty is $\sim$ 3$\%$ for each band.

The removal of stellar continuum contribution from H$\alpha$2 image is another critical part of data reduction. We used scaled R-band image as the stellar continuum, the scale factor was calculated using photometric counts ratios of 5 unsaturated field stars in both filter images. After the continuum was removed, the H$\alpha$2 image was flux calibrated following the conversion factor of the R band image and the effective transmissions of narrowband and R-band filters. In the present paper we don't consider the H$\alpha$ emission lost in the process of the continuum removal and the contribution of [N {\sc ii}] lines to the H$\alpha$ flux.

We also obtained optical spectra with Optomechanics Research, Inc. (OMR) Spectrograph using the same telescope on July 1, 2009. The spectra were taken with a 2$''$ width long slit along the major axis of the stellar bar, covering a wavelength range from 3800 to 8500\AA~in the observer's frame with the central wavelength at 6000\AA. We used the 200\AA/mm grating which resulted in a two-pixel resolution of 12\AA. The raw spectra were reduced with IRAF, and the preprocessing was similar to that of images mentioned above except the correction in the spatial and dispersion axis. After extracted, the spectra were made atmospheric absorption correction using the IRAF task TELLURIC. Then they were wavelength calibrated using the spectrum of He/Ar lamp, flux calibrated using the spectrum of standard star BD28+4211 observed at the same night.

%% Sec. 2.3
\subsection{Infrared Data}
We obtained the infrared images of NGC~7479 which were observed by Spitzer Space Telescope \citep{Werner04}. As a part of Mid-IR Hubble Atlas of Galaxies \citep{Fazio04a}, NGC7479 was imaged with the Infrared Array Camera \citep[IRAC;][]{Fazio04b} at 3.6, 4.5, 5.8 and 8.0 $\mu$m, as well as with the Multiband Imager and Photometer for Spitzer \citep[MIPS;][]{Rieke04} at 24, 70 and 160 $\mu$m. After we got the Basic Calibrated Data (BCD) which were generated through the Spitzer data reduction pipeline version s14.0.0, MOPEX (MOsaicker and Point source Extractor) version 18.1.5 was used to produce mosaicked images from individual BCD frames. The final images have spatial resolutions of 2.3$''$$\sim$2.6$''$, pixel sizes of 1$''$.2 for four bands of IRAC, and the spatial resolution of 6$''$, a pixel size of 2.5$''$ for MIPS 24m band. The pixel sizes of MIPS 70 and 160 $\mu$m are 4$''$ and 8$''$ respectively, along with very low spatial resolutions, so they are not included in our later study.

There are mainly three components in 8 $\mu$m band, they are the polycyclic aromatic hydrocarbon (PAH) emission, dust-continuum emission and stellar continuum. To obtain properties of the dust emission, the contribution of stellar continuum should be removed. We used a scaled IRAC 3.6 $\mu$m image as the stellar continuum with the assumption that the entire 3.6 $\mu$m band emission is from old stellar population. The scale factor of 0.232 \citep{Helou04} is adopted, which has also been used many times by other authors \citep[e.g.,][]{Cao08, Zhu08}. This coefficient was derived with the assumption that the IRAC 3.6 $\mu$m emission is all due to stars, and based on the Starburst99 synthesis model \citep{Leitherer99}. Occasional foreground stars located in the fields of NGC 7479 were almost removed by this technique, which suggests that this approach is fairly accurate in removing stellar emission. Although the 3.6 $\mu$m image is also somewhat contaminated by dust emission and red giant stars, particularly in the star forming regions, this component has only an impact of less than a few percent on the stellar continuum subtraction process \citep{Calzetti05}. Hereafter we marked the 8 $\mu$m dust emission in which stellar continuum has been removed as 8 $\mu$m(dust).

%%	The central region of NGC~7479 was also observed by the Infrared Spectrograph \citep[IRS;][]{Houck04} on Spitzer on 2003 December 17 and 2007 August 4 twice. The infrared spectrum are of four separated modules with low (R$\sim$60-130) and high (R$\sim$600) resolution from 5.2 to 38 $\mu$m. The reduction and analysis of the IRS spectrum has been made with CUBISM \citep{Smith07}.

%% Sec. 2.4
\subsection{Object Masking}
\label{objectmask}
In order to obtain accurate results, we removed the foreground bright field stars and background galaxies in images of optical B, V, R broadbands and two bands of IRAC. We made object masking following the method in \citet{Munoz09}. First, we used Sextractor \citep{Bertin96} to detect objects in each image, and then we selected the field stars and background galaxies using the parameters (FLUX and CLASS\_STAR) yielded from the first step. Finally, we replaced the selected sources with their nearest background values to mask them. As to the UV images, there are a mount of star formation regions located in them, it is hard to detect stars from these regions exactly, so the masking was not performed to these images. This reduction was also not made to 5.8$\mu$m, 8.0$\mu$m, 24$\mu$m images because the contribution from those masking objects could be negligible in mid-IR.

\begin{figure}
\center
\includegraphics[angle=0,scale=.8]{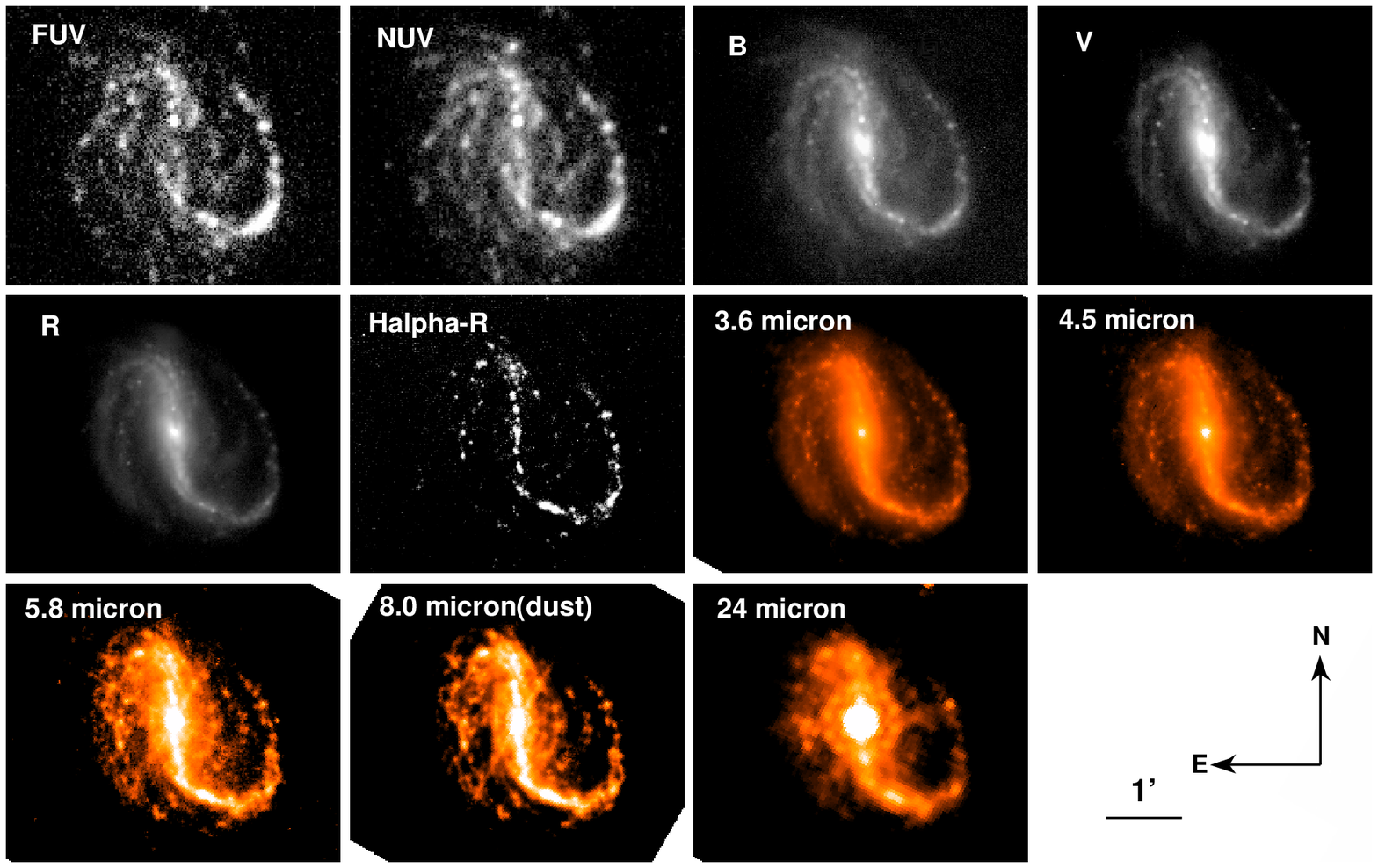}
\caption{The images of NGC7479 from UV to infrared. From left to right and top to bottom, they are FUV, NUV images from GALEX, B, V, R, H$\alpha$ images from Xinglong 2.16m telescope, 3.6, 4.5, 5.8, 8.0 and 24 $\mu$m from Spitzer. Note that H$\alpha$ image is subtracted by the stellar continuum using scaled R band image, 8 $\mu$m(dust) is the result of subtracting old stellar emission with 3.6 $\mu$m image. we have removed the bright field stars and background galaxies in the images of optical B, V, R broadbands and 3.6, 4.5 $\mu$m bands of IRAC (see Sec~\ref{objectmask}).
\label{fig1}}
\end{figure}

Figure~\ref{fig1} shows the final UV, optical, infrared images. From the figure, a strong stellar bar can be found in optical and near infrared images, and large asymmetry exists in UV and far infrared images.

%% Sec. 3
\section{Analysis \& Results}

%% Sec. 3.1 photometry
\subsection{Surface Photometry \& Radial Profile}
\label{photometry}
In order to study the global properties of NGC~7479, we used IRAF task ELLIPSE to make the surface photometry of multi-band images. The radial surface brightness profile, ellipticity ({\it e}) and position angle (PA) were included in photometric results of each band. The bar structure was obviously identified by following the method in \citet{Jogee04}. The characteristic bar signature is that {\it e} increases continuously untill reaching a maximum ${\it e}_{max}$, meanwhile PA remains a constant, and at the end of the bar, {\it e} falls rapidly and position angle changes correspondingly. ${\it e}_{max}$ was adopted as the ellipticity of the bar and the bar length was derived as the semi-major axis at which the maximum ellipticity was reached. We also found a slight jump in the bar region from the radial surface brightness profile. Based on R band image, we got that the length of the bar is about 8.3 kpc, the ellipticity 0.733, both are larger than their mean values in the local universe, $\sim$3.5 kpc and $\sim$0.6, respectively \citep{Marinova07, Gadotti10}. The length is much larger than the mean bar size of Sbc galaxies, which is 2.35 $\pm$ 1.22 kpc derived by \citet{Erwin05}. The ellipticity is also in the upper range of those in Sb/Sbc galaxies, which lie in the range 0.35--0.80 as derived by \citet{Marinova07}.

\begin{figure}
\center
\includegraphics[angle=0,scale=.8]{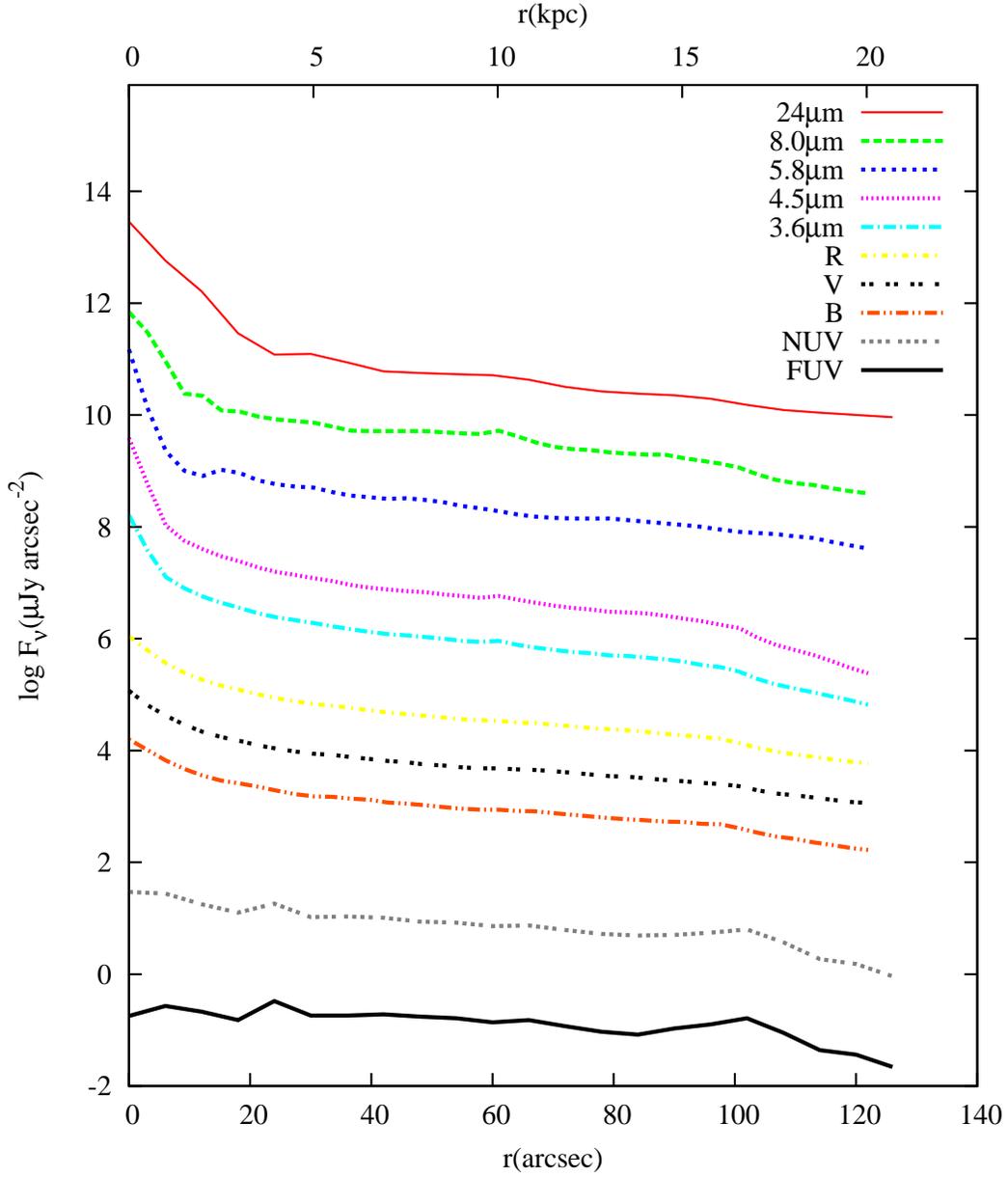}
\caption{Multi-wavelength surface brightness profiles for NGC 7479. From top to bottom, the profiles are shifted and arranged in the order of decreasing wavelength, FUV remains unchanged, NUV adds by 1, B adds by 2, and so on, 24$\mu$m adds by 9.
\label{fig2}}
\end{figure}

Because different band images have different spatial resolution as mentioned in Sec~\ref{data}, we did the surface photometry with two sets of radial profiles: one radial step of 3.05$''$ (for optical and IRAC images), the other of 6.1$''$ (for UV and 24 $\mu$m images). In order to get the same regions of different bands and compare with previous results easily, we used fixed centric position, ellipticity and position angle from the values provided in RC3 \citep{de91}. The result is shown in Figure~\ref{fig2}.
The properties of profiles mainly follow three behaviors: (1)mid-IR. In 24 $\mu$m band, there is a sharp inner cutoff at $\sim$20$''$, and there are also cutoffs at smaller radii in 8 $\mu$m and 5.8 $\mu$m images. Two factors may cause the high infrared brightness in the inner region: the first is that there are plenty of gas and dust reservoired in the central region, which are traced by 24 $\mu$m and 8 $\mu$m(dust) emission, respectively; the other might be that the emission is affected by the active galactic nucleus (AGN), especially for the 24 $\mu$m flux \citep[e.g.,][]{Bell05}. (2)UV. UV emission is mainly dominated by young stars. No strong peaks of UV profiles are found as those in infrared in the central region, this may be due to the strong dust attenuation. (3)optical \& near-IR. These profiles are relatively smooth and no obvious jumps can be found in the inner and outer disk. All the UV and optical photometric results have been corrected by the Galactic extinction with the assumption of the \citet{Cardelli89} extinction curve and R$_V$ = 3.1.

%% Sec. 3.2 Asymmetry & Concentration Index
\subsection{Morphological Asymmetry \& Concentration Index}
\begin{figure}
\center
\includegraphics[angle=0,scale=.8]{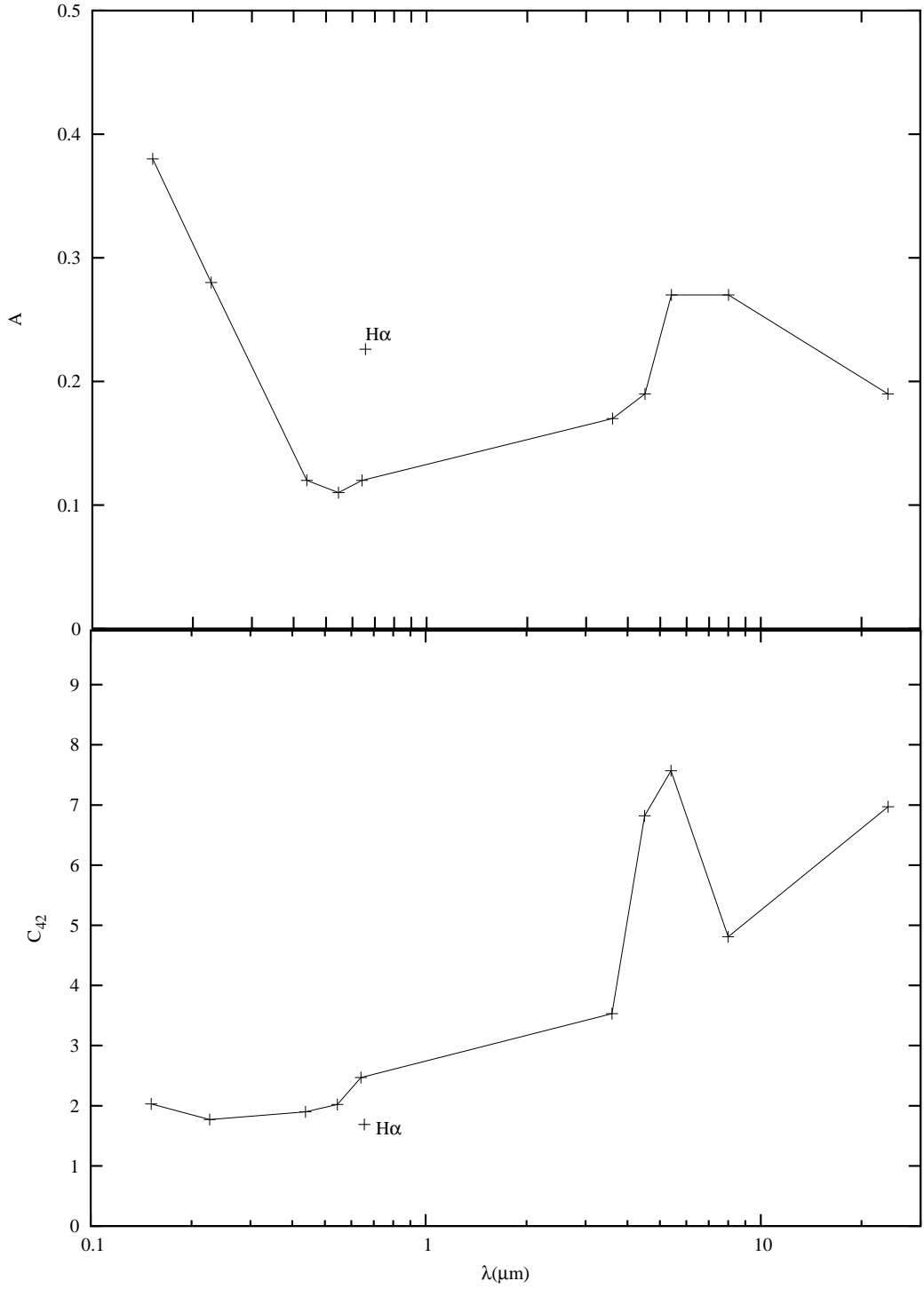}
\caption{{\it Top}: the concentration index C$_{42}$ at different wavelengths. {\it Bottom}: the Asymmetry as a function of wavelength. In both panels, the corresponding values of H$\alpha$ are marked out respectively.
\label{fig3}}
\end{figure}

The rotational asymmetry as a tool of describing properties of nearby and distant galaxies has been used in many studies \citep[e.g.,][]{Schade95, Conselice00}. The asymmetry is calculated following Equation (12) and (13) in \citet{Munoz09}, i.e., comparing the image completely inverted (i.e., the rotation angle is $180^\circ $) and the original counterpart:
\begin{equation}
A=\frac{1}{2}\left[\frac{\sum{\left| I_{180\degr}-I_0 \right|}}{\sum{\left| I_0 \right|}}-\frac{\frac{2}{\sqrt{\pi}}\sigma_{\mathrm{sky}}N_{\mathrm{pix}}\label{eq_sky_asym}}{\sum{\left| I_0 \right|}}\right],
\end{equation}
where $I_0$ and $I_{180\degr}$ are intensities of the original and rotated images, $\sigma_{\mathrm{sky}}$ is the sky noise, $N_{\mathrm{pix}}$ is the number of pixels within the aperture used to derive the asymmetry. The results are plotted in the top panel of Figure~\ref{fig3} as a function of wavelength. The asymmetry is highest in FUV, which is nearly 0.4, and then it drops to 0.11-0.12 in the optical wavelength. We got another maximal value 0.27 at 5.8 $\mu$m and 8 $\mu$m as we moved toward longer wavelength from optical to infrared. It is consistent with what we found from Figure~\ref{fig1}. The asymmetry of H$\alpha$ image is $\sim$0.226 a little smaller than the NUV value. The high asymmetry is probably due to the following reasons. The recent star formation regions traced by UV have a clumpy and extended spatial distribution and these regions sprawl out not only in the galactic center but also in the disk, which is not like middle and old aged stars traced by the optical and near infrared. The 5.8 $\mu$m and 8 $\mu$m emission are possibly due to 6.2, 7.7, 8.6 $\mu$m PAHs emission, although a large amount of dust is concentrated in the bulge and bar regions as the obvious dust lane along the bar is the proof, there are also a certain amount of dust in the arms and disk.

The concentration indexes are the ratios of two radii containing fixed fraction of the total galactic flux and they are widely used to identify structural properties of galaxies and infer the galactic morphological types \citep[e.g.,][]{Strateva01, Conselice03, Gil07}, as these indexes are relative parameters not affected by external factors. Here we adopt the index C$_{42}$ \citep{Kent85} defined to be
\begin{equation}
C_{42}=5log_{10}(r_{80}/r_{20}),
\end{equation}
where r$_{80}$ and r$_{20}$ are the radii along the semi-major axis containing 80\% and 20\% of the total luminosity. The bottom panel of Figure~\ref{fig3} shows the result along the wavelength, presenting a much different behavior. The index at UV is very small ($\sim$2) due to the contribution of the diffused young stars in the disk, and it is still small in optical wavelength. While in IR, C$_{42}$ jumps to maximal values 6.8 at 4.5 $\mu$m and 7.5 at 5.8 $\mu$m, and goes down at 8.0 $\mu$m for the spatial distribution of dust. Then it raises again at 24 $\mu$m due to its intense emission in the central region.

%% Sec. 3.3 HII regions
\subsection{Star Formation Properties of Different Structural Regions}
\label{regional-division}
\begin{figure}
\center
\includegraphics[angle=0,scale=.8]{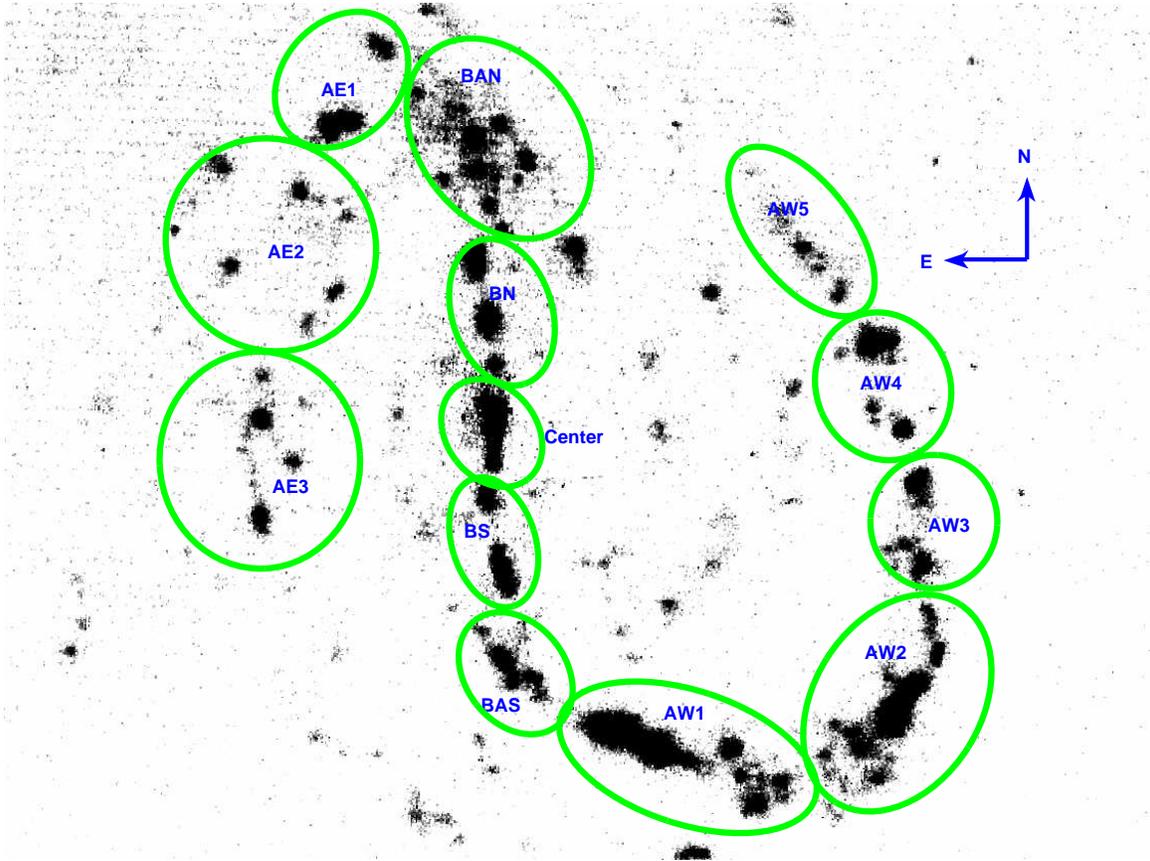}
\caption{The regional division on H$\alpha$ image. The ellipses overlapped in the image are regions we divided based on the H$\alpha$ image, marked with the corresponding labels. The bulge and nucleus region is plotted with red color. Although there are overlaps for some adjacent regions, it has little effect at the results (see Sec~\ref{regional-division}).
\label{fig4}}
\end{figure}

In the H$\alpha$ image (Figure~\ref{fig1}), we can find that most H {\sc ii} regions are located in the bar and two very asymmetric arms. These regions have been investigated in previous studies \citep[e.g.,][]{Rozas99, Zurita00}. While these studies are mainly focused on the global properties of the H {\sc ii} regions over the whole galaxy, from which the properties of star formation in different galactic structural regions can't be studied clearly. Therefore, in order to analyze properties of star formation regions situated in different structures, we marked 13 different regions along the bar and arms in Figure~\ref{fig4}. Because we focused on the entire properties of different regions not just on separated star formation clumps, several clumps are always contained in one marked region, which has the aperture of different radii and shape with each other. It should be noted that the results from this step just make sense on average in each region along with a certain degree of uncertainties. In Figure~\ref{fig4} all regions are marked with boundaries and labels on the H$\alpha$ image, including the southern and northern parts of the bar (noted as BS and BN), the bulge region (Center), three parts along the eastern arm, i.e., the weaker one (AE1--AE3), four parts along the other arm (AW1--AW4), and two parts at the interface between the bar and arms (BAN and BAS). The detailed information of each part is listed in Table 2.

%% Sec. 3.3.1 regional photometry & SED
\subsubsection{UV-to-Infrared SEDs}
\begin{figure}
\center
\includegraphics[angle=0,scale=.8]{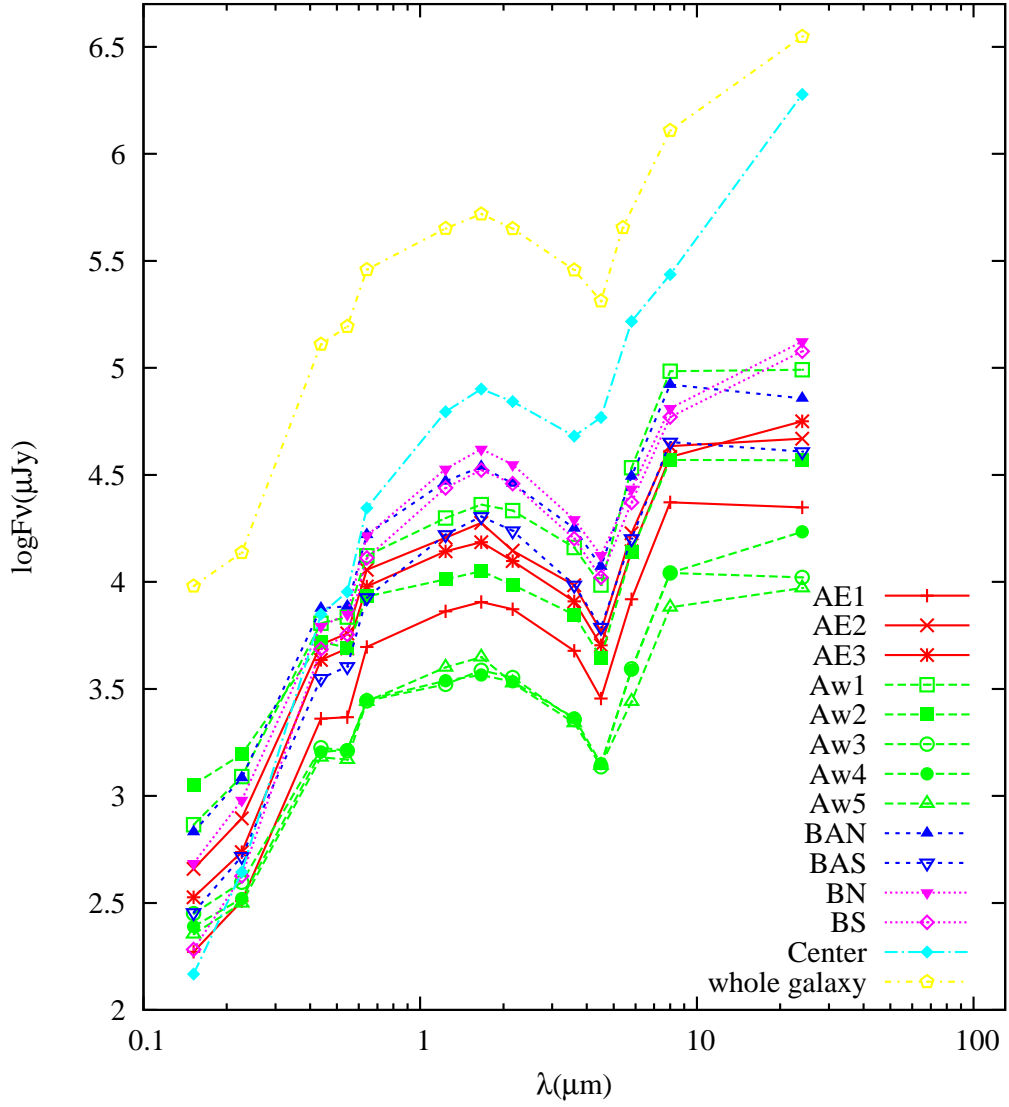}
\caption{Total SED of NGC 7479 and SEDs of individual H {\sc ii} regions marked in Sec~\ref{regional-division}. The SEDs from the same galactic structures are plotted with lines of the same color.
\label{fig5}}
\end{figure}

We made photometry of the entire galaxy and every marked region using the ELLIPSE task in IRAF on multi-wavelength images. The results are listed in Table~\ref{tabl2}, and they are also plotted in Figure~\ref{fig5} as the spectral energy distributions along the wavelength. To make the SEDs more accurately, we also added the photometric results from near-IR images of the J, H, and Ks Two Micron All Sky Survey (2MASS) bands \citep{Skrutskie06}.

For the entire galaxy, the fluxes vary by a factor of two orders of magnitude from 0.01Jy in the UV to above 3.5Jy in 24 $\mu$m. The optical fluxes are similar to the flux at 3.6 $\mu$m and they are dominated by the stellar emission. There is a steep slope from UV to optical wavelength, probably due to the Balmer jump or 4000$\AA$ break. It is clearly that the highest IR-flux and lowest UV emission lay at the galactic center in the thirteen structural regions. In previous studies, the nucleus of NGC 7479 was classified as Seyfert 1.9 \citep{Ho97b}, thus the emission from this region may be seriously affected by AGN, and it is hard to determine the proportion of the contribution from the AGN. Especially, the profiles of two regions in the bar also have higher IR emission than the regions in two spiral arms. For other SEDs, the profiles of regions from the same galactic structures are similar with each other, while all of their IR emission is much lower than that of the center.

% Sec. 3.3.2 Star formation rate
\subsubsection{SFR}
\label{SFR}
Star formation rates (SFRs) are very important to describe the activities of galactic star formation. Many tracers have been explored to estimate SFRs in different types of galaxies from UV to IR including broadband photometry and emission line luminosities \citep[e.g.,][]{Kennicutt98, Wu05, Zhu08, Kennicutt09, Calzetti10, Panuzzo10}.

In order to calculate SFRs more accurately, we used three methods obtained from previous studies. They are listed as follows:\\
\citet{Kennicutt09} (their Eq.[11] and Table [4]):
\begin{equation}
SFR_{H\alpha +24\mu m}(M_{\odot}yr^{-1})=7.9\times 10^{-42}[L(H\alpha)_{obs}+0.020L(24)](erg~s^{-1}) \label{eq_SFR1},
\end{equation}
\begin{equation}
SFR_{H\alpha +8\mu m}(M_{\odot}yr^{-1})=7.9\times 10^{-42}[L(H\alpha)_{obs}+0.010L(8)](erg~s^{-1}) \label{eq_SFR2}.
\end{equation}
They were derived on basis of the data from SINGS survey \citep{Kennicutt03} and \citet{Moustakas06}, and the calibration of \citet{Kennicutt98} with the assumption of solar abundances and the Salpeter IMF (0.1--100M$_{\odot}$). These methods are suggested to be applied reliably to individual H {\sc ii} regions and to galaxies as a whole \citep{Kennicutt09}.\\
\citet{Fisher09} (their eq.[6] \& [7]):
\begin{equation}
SFR_{FUV +24\mu m}(M_{\odot}yr^{-1})=2.21\times 10^{-43}[L(FUV)_{obs}+L(24)](erg~ s^{-1}) \label{eq_SFR3},
\end{equation}
which combines the 24$\mu$m IR emission from warm dust with the FUV emission from young stars. It was derived under the basis of star-forming regions in nearby galaxies from \citet{Calzetti07}, which luminosities were measured using elliptical apertures. This combination of FUV and 24$\mu$m is also suitable for the regions in NGC 7479.

\begin{figure}
\center
\includegraphics[angle=0,scale=.8]{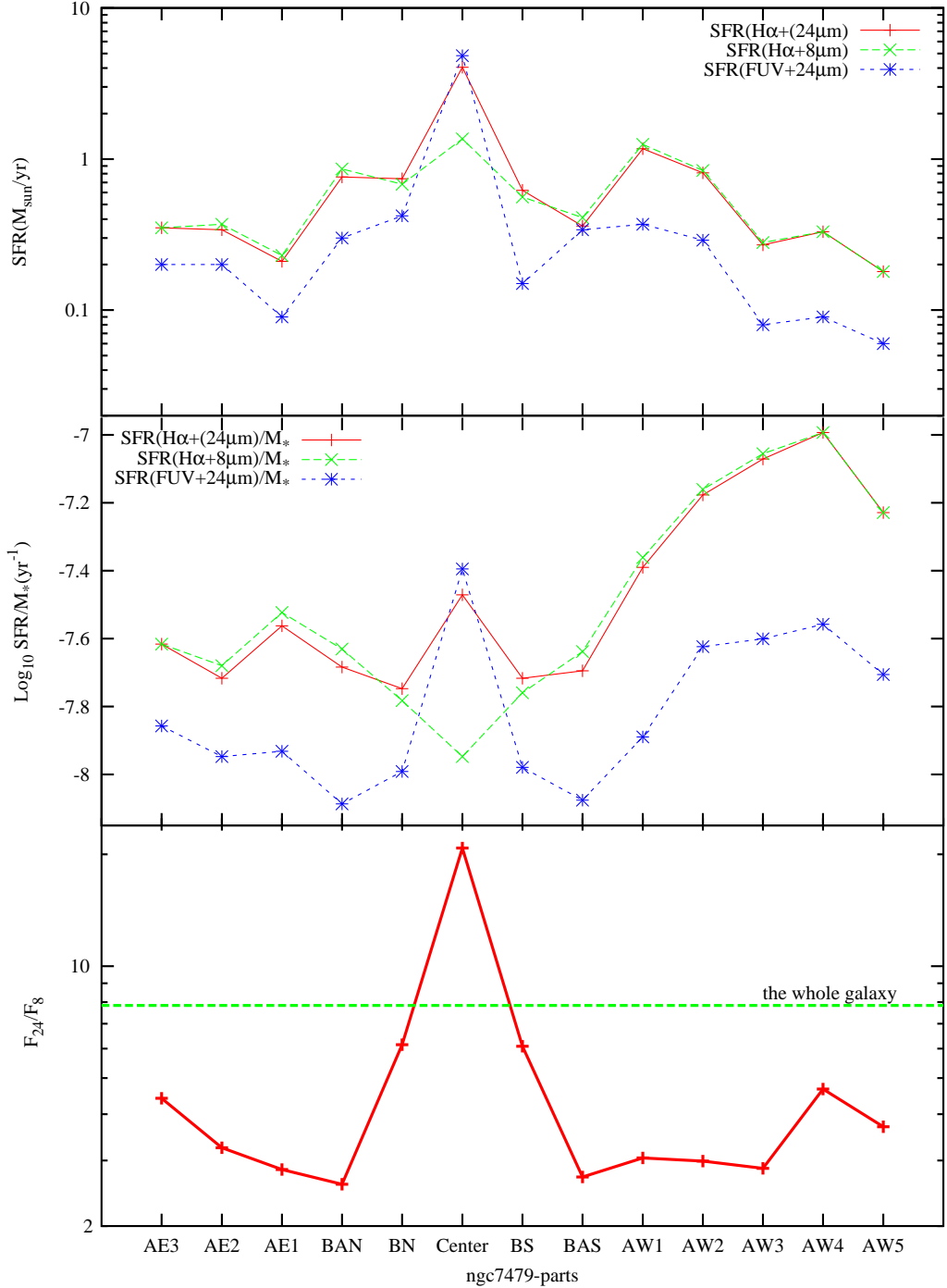}
\caption{{\it Top}: the star formation rate of each galactic region (marked in Figure~\ref{fig4}) calculated using different tracers. The SFRs from each method are arranged in the order of regional positions from the end of the eastern spiral arm to the western arm along the galaxy. In the central region, The SFRs calculated using the emission from 24$\mu$m are not reliable due to the effect of AGN. {\it Middle}: the specific star formation rate (SSFR, star formation rate [SFR] per unit galaxy stellar mass) of each region. {\it Bottom}: the flux ratio of 24$\mu$m and 8$\mu$m at each region. The flux of is 8$\mu$m emission from 8$\mu$m(dust) image.
\label{fig6}}
\end{figure}

In the top of Figure~\ref{fig6}, we plotted our results as positions in our galaxy. The SFRs derived from different equations show similar behaviors in most regions, while the value in the central region varies from $\sim$4.8 M$_{\odot} $yr$^{-1}$ calculated using UV + 24$\mu$m emission (Eq.~\ref{eq_SFR3}) to $\sim$1 M$_{\odot} $yr$^{-1}$ estimated using composite tracer of H$\alpha$ + 8$\mu$m emission (Eq.~\ref{eq_SFR2}). Values from the two formulas including 24 $\mu$m flux are larger than that from Eq.~\ref{eq_SFR2}, probably due to the effect of AGN. To further confirm this, we plotted VSG(very small grain)-to-PAH ratios characterized by F(24$\mu$m)/F(8$\mu$m) in the bottom of Figure~\ref{fig6}, because AGN activity is well correlated with these ratios \citep{Wu07}. In this plot, the central region shows the highest 24$\mu$m excess. Two barred regions also have higher F(24$\mu$m)/F(8$\mu$m) than the arm regions, it is likely that the F(24$\mu$m) of both barred regions is overestimated due to the contamination from the diffraction of the galactic nucleus in 24$\mu$m image. So the SFRs calculated using Eq.~\ref{eq_SFR1} and Eq.~\ref{eq_SFR3} in the three regions are probably higher than their real values. In the galactic disk, the SFRs from Eq.~\ref{eq_SFR3} are lower than those from Eq.~\ref{eq_SFR1} and Eq.~\ref{eq_SFR2}, likely due to the different initial mass function (IMF) used in these formulae. After our comparison, we adopt results from the composite tracer of H$\alpha$ and 8$\mu$m flux to do the next analysis. Except for the central region, star formation activity is mainly located at the strong arm and the end of the bar. The total SFRs are 0.95, 2.88, 1.24 and 1.36 M$_{\odot} $yr$^{-1}$ in the eastern arm, western arm, the bar and central region, respectively.

In order to describe explicitly how the recent star formation contribute to galaxy growth for different regions, we investigated the specific star formation rate (SSFR), which is defined as the SFR per unit stellar mass. Stellar masses are roughly determined with the luminosity of IRAC 3.6$\mu$m using the method from \citet{Zhu10} (their Eq. [2]):
\begin{equation}
Log_{10} {~M(M_{\odot})}=(-0.79\pm0.03)+(1.19\pm0.01)\times Log_{10} { ~\nu L_{\nu}{[3.6\mu m]}(L_{\odot})}\label{eq_mass}.
\end{equation}
It was deduced based on the sample cross-identified from the Spitzer SWIRE field and SDSS spectrographic survey.
The result is plotted in the middle of Figure~\ref{fig6}, where large difference is revealed in all marked regions. The west spiral arm shows greater values than other regions, while the SSFR arrives the lowest value in the central region.

%% Sec. 3.4 stellar population along the bar
\subsection{Stellar Population}
\begin{figure}
\center
\includegraphics[angle=0,scale=.8]{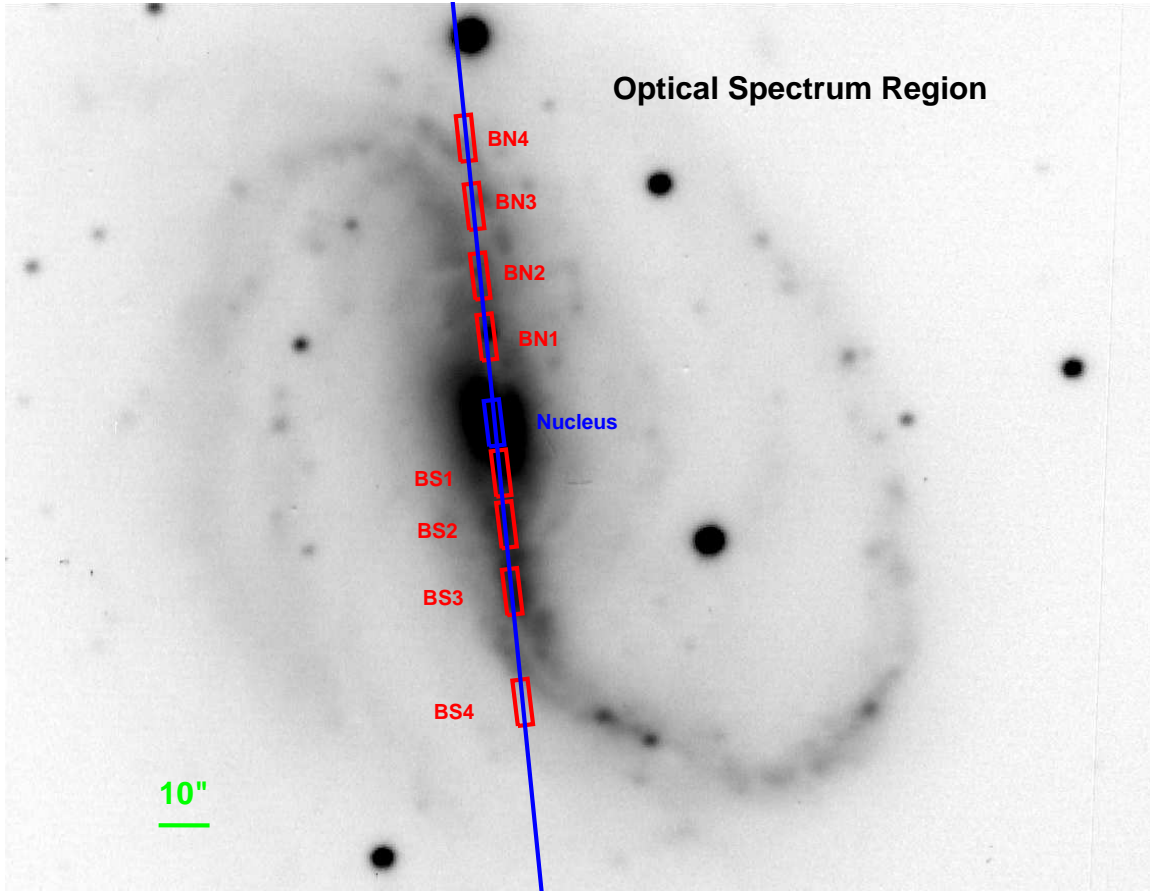}
\caption{Illustration of the optical spectrum long slit over-plotted on the raw R band image. The slit is signified by the blue line, the rectangles are the regions where spectrum are extracted in the bar (red, labeled as BN and BS) and nucleus (blue, labeled as Nuclues), the background used to subtract is located in the ends of the slit free from contamination of galaxies and stars. North is toward the up, east towards the left.
\label{fig7}}
\end{figure}

\begin{figure}
\center
\includegraphics[angle=0,scale=.75]{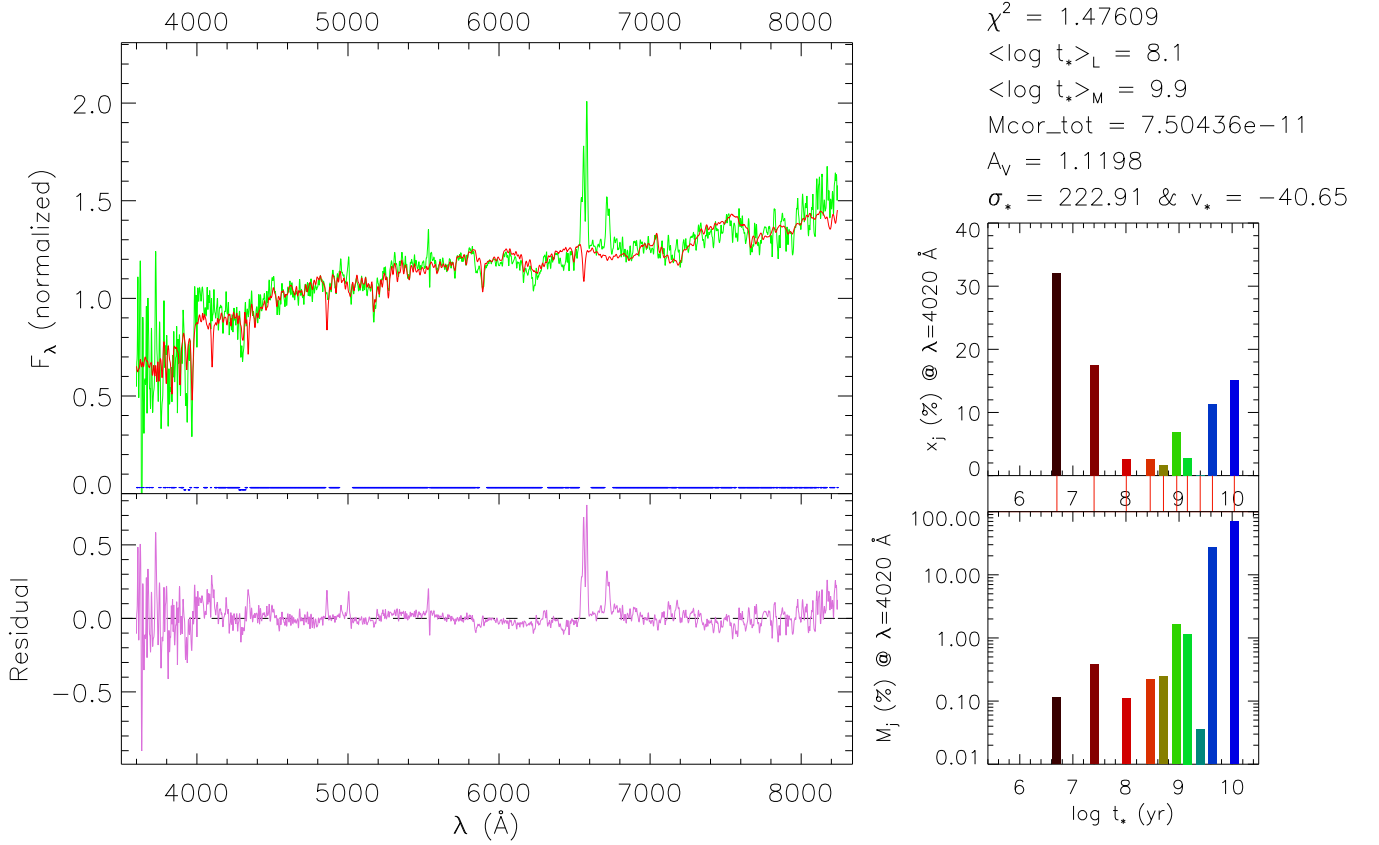}
\caption{Spectral synthesis of the nucleus of NGC 7479. {\it Left-top}: the observed spectrum (green), the model spectrum (red) and the error spectrum (blue) with the gaps where the region is not involved in the fitting. {\it Left-bottom}: the residual spectrum (purple). The spectrum in the left two panels are both normalized by the flux intensity at 4020\AA. {\it Right}: luminosity (top) and mass (bottom) weighted stellar population fractions x$_j$ and M$_j$ \citep{Cid05} as the function of ages of the stellar population templates. The basic results are also listed on the {\it right} panel.
\label{fig8}}
\end{figure}

We extracted optical spectra from 9 regions in the spectroscopic long slit along the stellar bar including the nucleus with the aperture of 9$''$(along the slit)$\times$2$''$(the width of the slit). These regions are shown in Figure~\ref{fig7}, all positions are determined with the H$\alpha$ emission peak, and are labeled as BN1--4 on the northern bar, BS1--4 on the southern bar, Nucleus at the galactic center. The spectral synthesis code STARLIGHT \citep{Cid05} is used to derive stellar population of different star formation regions along the bar. Previous studies have shown that this spectrum fitting code is a good tool to study the properties of various types of galaxies \citep[e.g.,][]{Meng10, Lara10, Martins10}. Our template spectra consist of simple stellar populations (SSPs) from \citet{BC03}. Assuming that the stellar populations have instantaneous burst star formation history, these SSP spectra are computed with ``Padov 1994" evolutionary tracks \citep{Alongi93, Bressan93, Fagotto94a, Fagotto94b, Girardi96} and \citet{Chabrier03} initial mass function (IMF). We used templates with ages of 0.005, 0.025, 0.1, 0.29, 0.5, 0.9, 1.4, 2.5, 4 and 10Gyr and metallicities of Z = 0.0001, 0.0004, 0.004, 0.008, 0.02 and 0.05.
Figure~\ref{fig8} shows the nuclear spectrum fitting result as an example. In this figure we obtained that the mean stellar age weighted by luminosity $\langle t_{*}\rangle _{L}$ \citep{Cid05} is $\sim$100Myr, and more than half of luminosity is dominated by populations younger than 100Myr. Besides young stellar populations, there is also an old population of $\sim$ 10 Gyr as the background population. We summarized our result of spectral synthesis in Table \ref{tabl3}. This result indicates that there were strong star formation activity on the bar $\sim$100Myr ago, especially at the two ends of the bar, where young stars account for a much higher percentage. Besides that, we can also learn that the stellar bar may have been $\sim$ 10Gyr old as old stars contribute to a large fraction in mass.

%% Sec. 4
\section{Discussion}

NGC 7479 is an interesting example with the strong stellar bar, and have likely suffered a minor merger because of some features mentioned above. The active star formation and distinctive morphology are evidences of the secular evolution driven by the stellar bar and maybe also by the minor merger event. In this section, we will discuss the effects of both drivers to our galaxy, respectively.

%% Sec. 4.1 Bar Driven Star Formation and Galactic Evolution
\subsection{Bar Driven Star Formation and Galactic Evolution}
\label{discussion_bar}
Based on what we have obtained above, the SFR in the bulge region of NGC7479 is $\sim$ 1 $M_{\odot}yr^{-1}$. The star formation activity in this region is likely related to the presence of the large-scale bar. Theoretically, stellar bars can concentrate a substantial fraction of disk gas in small galactocentric radii as a result of star formation \citep{KK04}. Observations also find that galaxies with stellar bars show higher central SFRs than unbarred ones on average \citep[e.g.,][]{Kormendy05, Fisher06, Shi06}, and the star formation in circumnuclear regions of spiral galaxies are more dependent on stellar bars than that in their outer disks \citep{Kennicutt98}. Using mid-IR color to trace star formation activity, \citet{Roussel01} also found that the color distributions of strong barred galaxies show 15 $\mu$m excesses, which are absent from weakly barred or unbarred galaxies. Using the specific SFR in the galactic central region of NGC 7479 (Figure~\ref{fig6}), we estimated that the growth time of the bulge at present SFR is about 50 $\sim$ 100 Myr, while \citet{Fisher09} derived that the median pseudobulge could have grown the present-day stellar mass in 8 Gyr, which is much longer than that of NGC 7479. Therefore, there is probably another factor affecting the evolution of this galaxy.

Besides the bulge, star formation activity is also located in the galactic bar of NGC 7479. \citet{Martin97} studied the morphological properties of H {\sc ii} regions in the bar of NGC 7479 using its H$\alpha$ image and found that the total SFR from H {\sc ii} regions in the bar excluding nucleus and circumnuclear region is about 0.4M$_{\odot} $yr$^{-1}$. \citet{LaineK99} also found the SFR is about 0.5M$_{\odot} $yr$^{-1}$ in the bar region, though this value is lower than what we have estimated using composite tracers. Previous studies also revealed that there is a large mount of gas distributing along the bar. Using CO (J=1 $\to $ 0) observations, \citet{LaineK99} found that CO emission in NGC 7479 is along the dust lane traversing the whole bar and nucleus. \citet{Huttemeister00} also found that a continuous gas distribution is all along the bar in NGC 7479 and the distributions of ${}^{12}CO$, ${}^{13}CO$, dust lanes and velocity gradient have complex relationships. The radio continuum emission at 21 cm in the galactic central region shows the similar behavior as the CO emission \citep{Laine98}. The distribution of star formation activity and gas in the bar may be a result of the concentration of gas inflowing from the disk under the influence of the bar, because the gravitational torque of the large-scale bar can make the gas lose angular momentum, and then drive it down inward through the stellar bar \citep{Athanassoula92, Sellwood93}.

In general, star formation locates in the circumnuclear regions and in the ends of bars, while bars are typically dominated by evolved stellar populations \citep{Gadotti06}. In the study of the barred spiral NGC 5383, \citet{Sheth00} found weak H$\alpha$ emission in the bar between the bar ends and the central region, despite the high gas column density in the bar dust lanes. Similarly, weak star formation in the bar takes place in NGC 1530, probably due to the inhibition of the shear and shock in the dust lanes \citep{Reynaud98}. However, as Figure~\ref{fig1} shows, obvious H$\alpha$ emission takes place in the bar of NGC 7479, and a majority of young stars are formed in nearly 100Myr (see Table~\ref{tabl3}). In addition, the SFR in the bar region is also comparable with those in the bulge and bar ends (Figure~\ref{fig6}). For that reason, the existence of star formation activity appears to be consistent with not only the galactic bar, but also another effect (see the next subsection).

How does the bar affect the evolution of NGC 7479? \citet{Quillen95} found that the gas inflow rate along the stellar bar is about 4--6 M$_{\odot}$yr$^{-1}$ in this galaxy, which is much larger than the SFR in the bar region we derived (1.24 M$_{\odot}$yr$^{-1}$). Therefore, before turning into stars a large percent of gas will drop into the galactic central region and make the bulge mass addition \citep{LaineK99}. In this case, this galaxy will possibly evolve toward an earlier galaxy type, although there is a doubt that a late-type galaxy maybe not evolve into an early type via bar-induced gas inflow \citep{Sheth05}. Therefore, NGC 7479 is likely just in a transitional stage.

In order to explore the bar-driven secular evolution, \citet{Jogee05} classified 10 different barred galaxies into three evolutionary stages based on the dynamical properties and dust morphology: (i) Type I Non Starburst, the early stages of bar-driven inflow, where large amounts of gas are along the bar and there is low star formation efficiency in the inner few kiloparsecs; (ii) Type II Non Starburst, the later stages of bar-driven inflow where most molecular gas is concentrated in inner kpc and star formation only occurs in regions with high gas densities; subsequently, (iii) Circurmnuclear Starburst, where large fraction of molecular gas is concentrated in galactic central region, exists intense star formation activity. Comparing our results with this scenario suggests that NGC 7479 may be in Type I Non Starburst evolutionary phase based on the properties of star formation and molecular gas in this galaxy. NGC 7479 is similar to NGC 4569, which is the prototype of Type I Non Starburst in the work of \citet{Jogee05}. With a sample of barred galaxies, \citet{Verley07} also put forward a scenarios of bar-driven secular evolutionary sequence using properties of H$\alpha$ emission, in which they defined four main groups, i.e., Group E, a strong central peak in the H$\alpha$ emission, no H$\alpha$ emission in bar; Group G, H$\alpha$ emission in the bar; Group EG, a transition between the E and G groups; Group F, with less gas, smoother morphology and no central emission spot in H$\alpha$, and argued that the evolutionary sequence is G $\to $ EG $\to $ E $\to $ F. In their studies, NGC 7479 is classified as Group G. Similar results are also found by \citet{Martin97}, our galaxy suggested to be in the middle transitional type in their three stages of evolutionary sequence.

 Thus, as performed by above authors, NGC 7479 is in a temporal transitional phase. Under the effect of its stellar bar, most of molecular gas will be concentrated in the inner kpc of the galactic center, starburst can be triggered in circumnuclear region when most of the gas exceeds a critical density, and pseudobulges is expected to be built in the inner kpc. Then NGC 7479 may become a galaxy with a earlier Hubble type. On the other hand, the increase of the infrared luminosity may also make our galaxy a luminous infrared galaxy (LIRG), this may explain why there are some LIRGs are isolated barred galaxies but not interacting systems \citep[e.g.,][]{Wang06}.

%% Sec. 4.2 Minor Merger Event in NGC7479
\subsection{Minor Merger Event in NGC7479}
Minor merger event is a common phenomenon in the Universe. According to statistics, each field spiral galaxy has more than one companions on average \citep{Zaritsky97}. Thus, the interactions between galaxies and their companions are believed to be very common \citep{Ostriker75}. Several properties of NGC 7479 indicate the presence of a recent minor merger. First of all, as we mentioned in Sec~\ref{discussion_bar}, star formation activity in the bar and bulge of our galaxy is stronger than most common barred galaxies. The systems in minor mergers have higher SFR than those found in normal H {\sc ii} regions of spiral galaxies \citep{Ferreiro08}. Minor mergers also likely play a role in enhancing the infrared luminosity for some low-luminosity galaxies \citep{Kartaltepe10}. These features are likely caused by a large amount of gas which was driven into host galactic disks when they swallowed gas-rich satellites, and the gas may also be transported into the galactic inner regions due to bars.

Second, in numerical simulations, vertical impact of small companions can cause asymmetries to gas-rich barred galaxies \citep{Berentzen03}. \citet{Walker96} found that minor merger can form significant disk asymmetry, visible for at least 1 Gyr in their simulations. Similarly, the simulations of \citet{Bournaud05} indicate that minor mergers in some conditions can create a strong asymmetry when the galaxy appears to be ``isolated'', once the most obvious merger related features have disappeared. Through comparing the distribution of different compositions in Figure~\ref{fig1}, we found that star formation regions of NGC 7479 are largely concentrated in the strong arm, and PAH emission is mostly distributed along the stellar bar and the strong spiral arm. Probably because of this reason, the asymmetry of NGC 7479 is much higher in UV, mid-infrared, as well as H$\alpha$ band (A $\sim$ 0.3), than that in the optical and near-infrared bands.

Third, minor mergers are probably related to the triggering of stellar bars \citep{Shumakova05, Emilio08}. Since active star formation exists in the bar of NGC 7479, the intense gas concentration and young stars in the stellar bar are probably driven by the minor merger event in some degree besides the stellar bar. The large percentage of burst population in the bar of NGC 7479 (Table~\ref{tabl3}) are evidently formed in recent hundred Myr. Furthermore, as argued by \citet{Martin00} there is strong evidence that the present morphology of NGC 7479 was acquired following a merging event with a small galaxy about 300 Myr ago.
In addition, the radio continuum jet in NGC 7479 found by \citet{Laine08} is another signal of recent perturbation by a minor merger in this galaxy, and is similar with the cause of jets in NGC 1097 \citep{Higdon03}. Through the comparison between the simulation model of minor merger and observation, \citet{LaineH99} suggested that NGC 7479 is the result of minor merger which is still under process, and the remnant of the satellite galaxy likely locates within the bar. Besides the minor merger, there are likely other processes to contribute large amounts of gas to the evolution of NGC 7479. For example, galaxy tidal encounters \citep{Li08} and external gas accretion \citep{Bournaud02} that can also result in material infalling; in the simulation of \citet{Sellwood99}, a tidal encounter or the accreted low angular momentum material could also strengthen the bar.

Although morphological properties of NGC 7479 are consistent with characteristics of minor merger, no companions have been found in the field of this galaxy \citep{Laine98, Saraiva03}. To explore whether any remnant of merging companions is still existing, we compared 0.2--12.0 keV X-ray images from XMM-Newton telescope, high-resolution images from HST Proposal 6266 with WFPC2 and the F814W filter. Since ultraluminous compact X-ray sources are probably dwarf companion galaxy nuclei \citep{Makishima00, Wu02}, and merger remnants if still exist are also likely located in compact optical sources, we try to find possible candidates from these sources.

\begin{figure}
\center
\includegraphics[angle=0,scale=0.40]{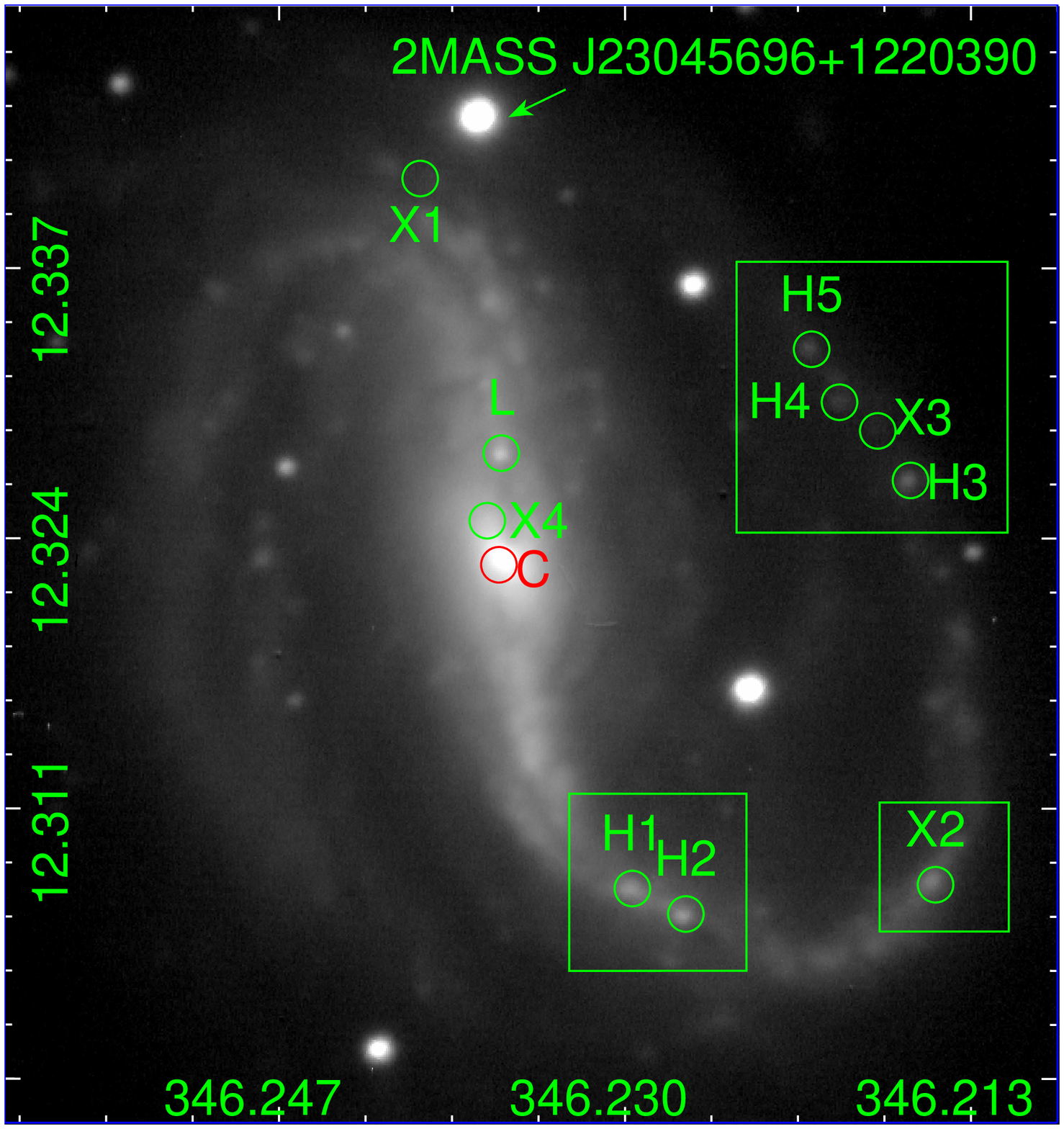}
\includegraphics[angle=0,scale=0.40]{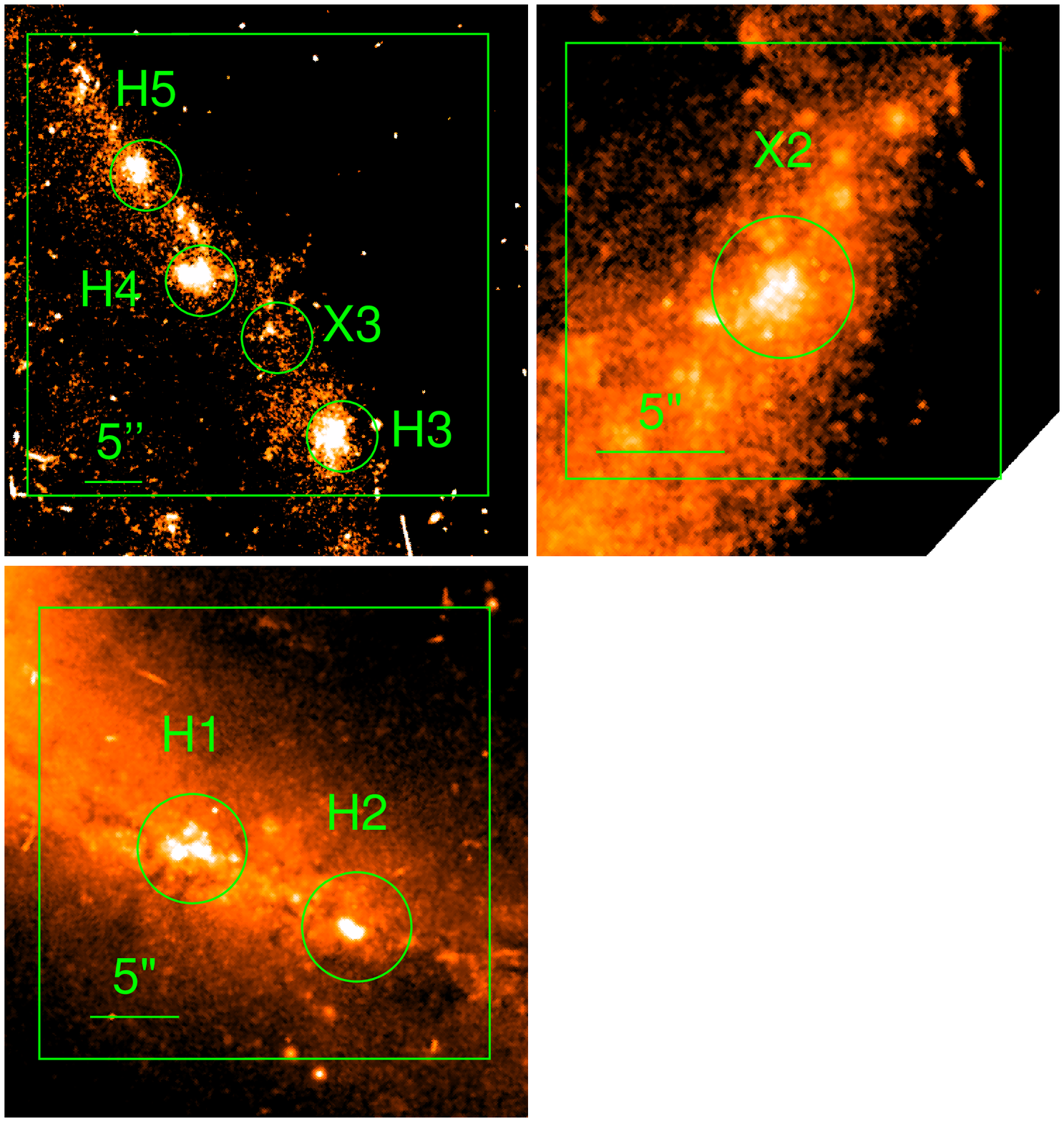}
\caption{Potential remains of minor merger. 10 potential positions are marked with green circles, 4 candidates(X1-X4) are identified with images from XMM-Newton telescope, 5 candidates(H1-H4) are identified used hight-resolution images from HST, The target L is estimated by \citet{LaineH99}. C(red circle) is the host galactic nucleus, which also has strong X-ray emission and can be detected by XMM-Newton. The {\it left} panel is the image from Xinglong 2.16m Telescope with R band, the radii of the circles are 3$''$. The {\it right} panel lists the zoomed images from HST with WFPC2 and the F814W filter, the radii of the circles are 2$''$.
\label{fig9}}
\end{figure}

 In Figure~\ref{fig9} we marked 10 candidates, among which there are 4 strong X-ray emission peaks (labeled as X1-X4) detected in X-ray image and 5 intensive optical regions (labeled as O1-O5) in HST image, the last one is a potential target pointed by \citet{LaineH99} (marked with L). Besides those, the nucleus of the host galaxy (marked as C) can also be clearly detected by XMM images. Target X4 is the nearest ($\sim$1kpc) X-ray emission peak to the galactic nucleus C, and it lies in the most obscured region, which may be the reason it can't be identified in optical wavelength, and distinguished from the nucleus. Therefore, we could not do the further investigations for this source. Target X1 located in the north end of the bar, has the most X-ray emission flux in the five peaks (assuming they are at the same distance), but it may be the foreground star 2MASS J23045696+1220390 if we consider the error in pointing (the mean error of 8$''$) and resolution (spatial resolution 6$''$ FWHM) of XMM-Newton telescope. Target X2 and X3 are both located in the west strong spiral arm. X2 can be detected obviously from the high-resolution optical image from HST. In the position X3, there is an optical counterpart with very weak emission, which is only $\sim$6$''$ from the compact source H4, also not far from two faint sources H3 and H5. H1 and H2 are located on the strong arm and near the stellar bar, especially H2 can not be resolved even in the images of HST. The potential target L lies on the north of the galactic nucleus in the bar, where there is little X-ray emission but an bright optical counterpart. The detailed information is listed in Table~\ref{tabl4}, the coordinates of X-ray candidates are from XMM-Newton detection, and their luminosity (L$_X$) are including the emission from 0.2keV to 12keV.

\begin{figure}
\center
\includegraphics[angle=0,scale=.8]{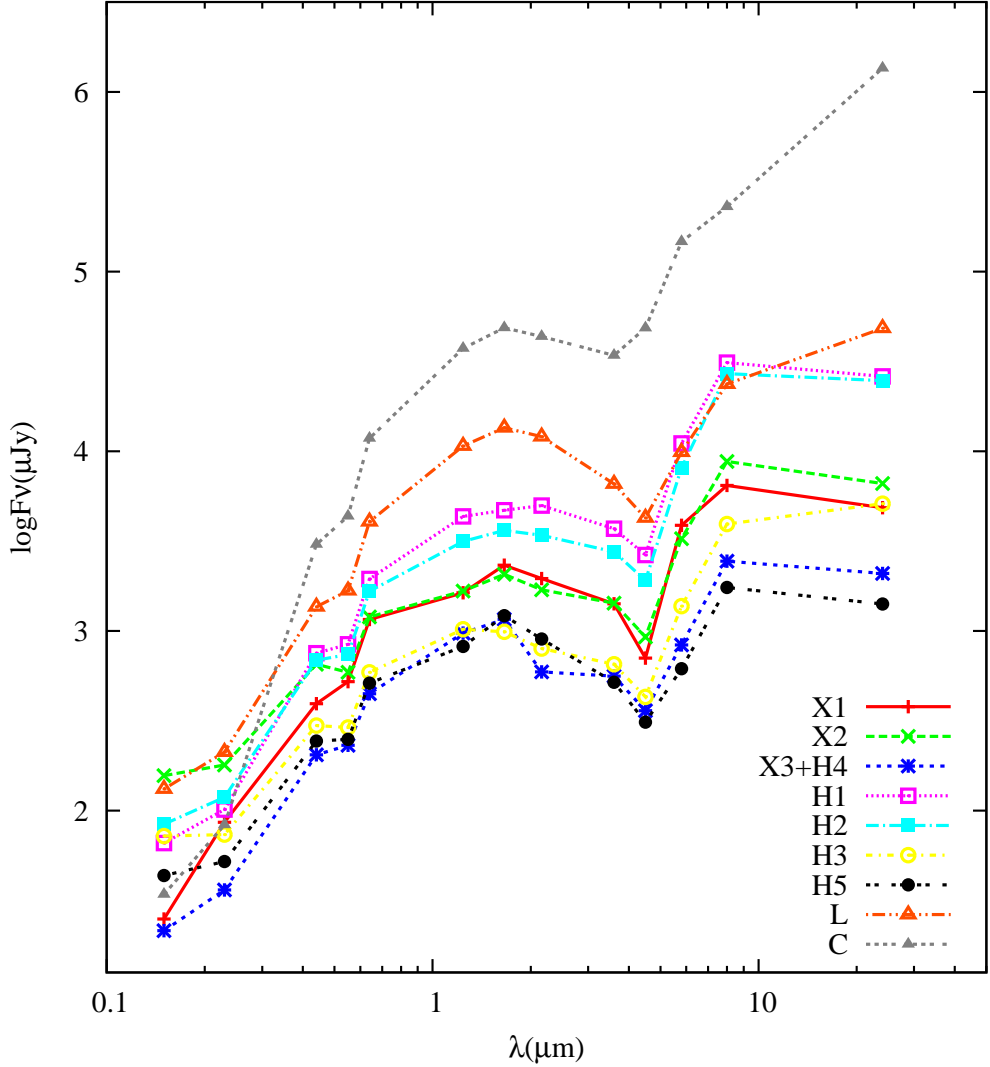}
\caption{{\it Left}: spectral energy distribution (SED) of each candidate of minor merger remnant compared with that of the nucleus of NGC 7479. The labels are the same with Figure~\ref{fig9}. Due to the low spatial resolution, X3 and H4 can not be taken apart in some images, we plot them together using their sum flux. All fluxes are from aperture photometry with radius of 6$''$.
\label{fig10}}
\end{figure}

For further investigations of properties of candidates, we made aperture photometry with radius of 6$''$ to get their SEDs (Figure~\ref{fig10}). Due to the low spatial resolution, we used the sum flux of X3 and H4 to plot their SED. We also obtain the VSG-to-PAH ratio using F(24$\mu$m)/F(8$\mu$m) and surface SFR using Eq.~\ref{eq_SFR2} with small photometric apertures (radii of $\sim$2$''$) (see Table~\ref{tabl4}). We found 3 sources have the positive ratios and 4 sources have surface SFR higher than 0.3 M$_{\odot} $yr$^{-1}$ kpc$^{-2}$. By comparing these plots, we found that sources L and H3 have similar SEDs with the nucleus C, and they also have higher VSG-to-PAH ratios, which indicates they are more likely in the criteria region of AGNs \citep{Wu07}. Thus, under the consideration of the two aspects, we suggest that the sources L and H3 are more likely remnants of the minor merger event.

%% Sec. 5 SUMMARY
\section{Summary}
In this paper, we present a multi-wavelength study of properties of the isolated barred galaxy NGC 7479 with ground-based optical images and spectra combining with GALEX UV and Spitzer IR data. Our results are as follows.

1. We took the surface photometry from far-UV, optical to mid-IR, and obtained radial profile of each band, and found that these profiles show different behaviors in different wavelengths. Based on R band image, we derived that the stellar bar in NGC 7479 is 8.3 kpc in length, 0.733 in ellipticity.

2. Calculating the morphological asymmetry and concentration index of NGC 7479, we found that the asymmetry of this galaxy is high in UV and mid-IR as well as the H$\alpha$ narrow band, while the concentration index is much small in UV and locates peak values in PAH emissions.

3. The properties of star formation regions in NGC 7479 are analyzed using several tracers. Intense star formation activity is found to take place on the bar and one strong spiral arm, which also has the highest special SFRs in the disk. The star formation activity on our stellar bar is comparable with those in the central region and in the ends of the bar, while the growth time of our galactic bulge at present-day SFR is also shorter than the typical value of Gyrs derived by \citet{Fisher09}

4. With the help of stellar synthesis code STARLIGHT, we found that strong star formation took place in the bar about 100Myr ago, and the stellar bar may have been $\sim$ 10Gyr old as old stars contribute to a large fraction in mass.

5. By comparing our results with previous studies, we suggested that our galaxy is in a temporal and transitional phase of the secular evolution. Under the effect of its galactic bar, NGC 7479 may become an earlier type galaxy or a LIRG finally.

6. The properties of active star formation and distinctive morphology in NGC 7479 also indicate that a minor merger event likely happened recently. If so, ten possible remnants of merger companions are measured, among which two candidates have been found.

In order to understand the roles of large-scale bars and minor merger to galaxy secular evolution, studies based on large samples are necessary. In the next step work, we will make further researches using our galaxy sample in the following series papers.

\begin{acknowledgements}
We are grateful to thank the anonymous referee, for his/her very careful review of this paper, much constructive advice, and very useful suggestions. We thank Juan Carlos Munoz for his generous help and assistance throughout the process of data masks. We acknowledge Jun-Cong Liu for the modification of the language in this paper. This project is supported by NSFC grant 10833006, 10773014, 10978014 and the 973 Program grant 2007CB815406.

This work was partially Supported by the Open Project Program of the Key Laboratory of Optical Astronomy, National Astronomical Observatories, Chinese Academy of Sciences. This work is based on observations made with the {\it Spitzer Space Telescope} and with the {\it Galaxy Evolution Explorer}. The {\it Spitzer Space Telescope} is operated by the Jet Propulsion Laboratory, California Institute of Technology, under NASA contract 1407. GALEX is operated for NASA by the California Institute of Technology under NASA contract NAS5-98034. We are also grateful to the University of Massachusetts and NASA/IPAC support for supplying 2MASS data. This work is based in part on observations made with the NASA/ESA {\it Hubble Space Telescope}, obtained from the data archive at the Space Telescope Science Institute. STScI is operated by the Association of Universities for Research in Astronomy, Inc. under NASA contract NAS 5-26555. This research is partially based on observations obtained with {\it XMM-Newton}, an ESA science mission with instruments and contributions directly funded by ESA Member States and NASA, and we are also grateful to the X-ray image observer Dr Kazushi Iwasawa. This research has also made use of the NASA/IPAC Extragalactic Database (NED) which is operated by the Jet Propulsion Laboratory, California Institute of Technology, under contract with the National Aeronautics and Space Administration.

\end{acknowledgements}

%% Figures (1-9)

% --- Table 1 ---

\begin{deluxetable}{cccccccc}

\tabletypesize{\scriptsize}
\tablecaption{Observation Log}
\tablewidth{0pt}
\tablehead{
\multicolumn{1}{c}{   } & \multicolumn{1}{c}{   } & \multicolumn{1}{c}{   } & \multicolumn{1}{c}{$\lambda _{eff}$} & \multicolumn{1}{c}{Exptime} & \multicolumn{1}{c}{Obs-Date} & \multicolumn{1}{c}{FWHM} & \multicolumn{1}{c}{Pixel size}\\
\multicolumn{1}{c}{Band} & \multicolumn{1}{c}{Telescope} &\multicolumn{1}{c}{Instrument} & \multicolumn{1}{c}{($\mu$m)} & \multicolumn{1}{c}{(s)} & \multicolumn{1}{c}{(UT)} & \multicolumn{1}{c}{($''$)} & \multicolumn{1}{c}{($''$/pix)}
}
\startdata
\multicolumn{8}{c}{\underline{Archival Observations}} \\
%\cutinhead{GALEX}
FUV & GALEX & & 0.1516 & 1606.05 & 2004-10-02 & 6.0 & 1.500 \\
NUV & GALEX & & 0.2267 & 1606.05 & 2004-10-02 & 6.0 & 1.500 \\
%\cutinhead{Spitzer}
Mid-IR & Spitzer & IRAC & 3.6/4.5/5.8/8.0 & 10$\times $26.8 & 2004-07-05 & 2.3-2.6 & 1.220  \\
Mid-IR & Spitzer & MIPS & 24 & 216$\times $2.62 & 2005-06-27& 6.0 & 2.500 \\
\multicolumn{8}{c}{\underline{Own Observations}} \\
%\cutinhead{Xinglong 2.16m}
B & Xinglong 2.16m & BFOSC & 0.438 & 1200 & 2009-09-12 & 2.8 & 0.305 \\
V & Xinglong 2.16m & BFOSC & 0.545 & 1100 & 2009-09-12 & 2.5 & 0.305 \\
R & Xinglong 2.16m & BFOSC & 0.641 & 600 & 2009-09-12 & 2.2 & 0.305 \\
H$\alpha$2 & Xinglong 2.16m & BFOSC & 0.661 & 3000 & 2009-09-12 & 2.0 & 0.305 \\
Spectrum & Xinglong 2.16m & OMR & 0.38--0.85 & 2$\times$3600 & 2009-07-02 & ... & ... \\
\enddata
\label{tabl1}
\end{deluxetable}

\clearpage
% --- Table 2 ---
\begin{deluxetable}{lllllllllllllllll}
\tabletypesize{\tiny}
\rotate
\tablecaption{Properties of the H {\sc ii} regions}
\tablewidth{0pt}
\tablehead{
\colhead{Regions} & \colhead{RA} & \colhead{DEC} & \colhead{Major r} & \colhead{Minor r} & \colhead{PA} & \colhead{FUV} & \colhead{NUV} & \colhead{B} & \colhead{V} & \colhead{R} & \colhead{3.6$\mu$m} & \colhead{4.5$\mu$m} & \colhead{5.8$\mu$m} & \colhead{8.0$\mu$m} & \colhead{24$\mu$m} & \colhead{H$\alpha$}\\
\colhead{name} & \colhead{J2000.0} & \colhead{J2000.0} & \colhead{arcsecond} & \colhead{arcsecond} & \colhead{degree} & \colhead{mJy} & \colhead{mJy} & \colhead{mJy} & \colhead{mJy} & \colhead{mJy} & \colhead{mJy} & \colhead{mJy} & \colhead{mJy} & \colhead{mJy} & \colhead{mJy} & \colhead{10$^{40}$erg s$^{-1}$}\\
\colhead{(1)} & \colhead{(2)} & \colhead{(3)} & \colhead{(4)} & \colhead{(5)} & \colhead{(6)} & \colhead{(7)} & \colhead{(8)} & \colhead{(9)} & \colhead{(10)} & \colhead{(11)} & \colhead{(12)} & \colhead{(13)} & \colhead{(14)} & \colhead{(15)} & \colhead{(16)} & \colhead{(17)}
}
\startdata
Center & 23:04:56.6 & +12:19:21.12 & 10.9 & 8.3 & 37.9 & 0.15 & 0.44 & 7.05 & 8.99 & 22.13 & 48.01 & 58.59 & 164.87 & 273.32 & 1894.25 & 7.79 \\
BN & 23:04:56.6 & +12:19:43.05 & 13.8 & 9.2 & 17.3 & 0.48 & 0.96 & 6.21 & 7.05 & 16.44 & 19.58 & 13.29 & 27.20 & 64.73 & 132.78 & 6.34 \\
BS & 23:04:56.5 & +12:19:00.23 & 12.3 & 7.7 & 17.3 & 0.19 & 0.42 & 4.85 & 5.68 & 12.96 & 15.94 & 10.48 & 23.48 & 58.95 & 119.63 & 5.12 \\
BAN & 23:04:56.7 & +12:20:13.61 & 20.3 & 14.9 & 37.3 & 0.68 & 1.21 & 7.52 & 7.69 & 16.62 & 17.77 & 11.78 & 31.09 & 83.49 & 72.07 & 7.97 \\
BAS & 23:04:56.3 & +12:18:37.17 & 12.3 & 9.2 & 37.9 & 0.29 & 0.52 & 3.54 & 4.02 & 8.45 & 9.66 & 6.14 & 15.95 & 45.03 & 10.65 & 3.68 \\
AE1 & 23:04:58.5 & +12:20:24.48 & 13.9 & 10.8 & -42.7 & 0.19 & 0.32 & 2.30 & 2.33 & 4.96 & 4.76 & 2.85 & 8.30 & 23.55 & 22.30 & 2.12 \\
AE2 & 23:04:59.5 & +12:19:54.73 & 20.0 & 19.0 & 37.9 & 0.46 & 0.79 & 5.09 & 5.74 & 11.39 & 9.64 & 5.88 & 16.88 & 43.09 & 46.73 & 3.23 \\
AE3 & 23:04:59.6 & +12:19:15.35 & 20.0 & 18.4 & 2.9 & 0.34 & 0.55 & 4.32 & 4.89 & 9.52 & 8.11 & 5.05 & 13.90 & 38.33 & 56.33 & 3.13 \\
AW1 & 23:04:54.3 & +12:18:19.07 & 24.6 & 12.3 & 71.9 & 0.73 & 1.23 & 6.41 & 6.86 & 13.24 & 14.47 & 9.68 & 34.17 & 96.42 & 97.97 & 12.46 \\
AW2 & 23:04:51.6 & +12:18:29.46 & 21.6 & 15.4 & -32.1 & 1.13 & 1.56 & 5.24 & 4.93 & 8.55 & 7.03 & 4.42 & 13.83 & 37.16 & 36.98 & 9.38 \\
AW3 & 23:04:51.3 & +12:19:02.55 & 12.3 & 11.7 & -3.1 & 0.28 & 0.40 & 1.68 & 1.63 & 2.78 & 2.28 & 1.37 & 3.92 & 11.01 & 10.50 & 3.12 \\
AW4 & 23:04:51.7 & +12:19:29.02 & 13.9 & 12.3 & 22.9 & 0.25 & 0.33 & 1.60 & 1.64 & 2.81 & 2.31 & 1.41 & 3.87 & 11.01 & 17.15 & 3.76 \\
AW5 & 23:04:52.8 & +12:19:57.60 & 18.5 & 9.2  & 37.9 & 0.23 & 0.32 & 1.52 & 1.48 & 2.76 & 2.19 & 1.39 & 2.76 & 7.60 & 9.37 & 2.02 \\
Entire & 23:04:56.6 & +12:19:22.40 & 123.0 & 93.0 & 22.0 & 9.56 & 13.70 & 129.01 & 156.09 & 288.11 & 286.33 & 205.20 & 452.25 & 1285.60 & 3531.18 & 110.17 \\
\enddata
\tablecomments{Columns: \\
(1): name of galactic region. \\
(2)-(3): right ascension and declination in J2000.0. \\
(4)-(6): the size and position angle of each region. The angles run from 0 in the north, counterclockwise. \\
(7)-(16): the photometric result of each wavelength-band, in unit of mJy. \\
(17): the photometric result of H$\alpha$ narrow-band, including the emission line [N {\sc ii}], in unit of 10$^{40}$erg s$^{-1}$}.
\label{tabl2}
\end{deluxetable}

\begin{deluxetable}{ccccccccc}

\tabletypesize{\scriptsize}
\tablecaption{Results of Spectral Synthesis}
\tablewidth{0pt}
\tablehead{
\colhead{Region} & \colhead{Log $\langle t_{L}\rangle$} & \colhead{Log $\langle t_{M}\rangle$} & \colhead{Burst\%} & \colhead{Young\%} & \colhead{Median\%} & \colhead{Old\%} & \colhead{A$_{v}$} & \colhead{$\chi^2$} \\
\colhead{(1)} & \colhead{(2)} & \colhead{(3)} & \colhead{(4)} & \colhead{(5)} & \colhead{(6)} & \colhead{(7)} & \colhead{(8)} & \colhead{(9)}
}
\startdata
BN1 & 7.1 & 10.0 & 88.41 & 0.00 & 0.00 & 11.59 & 1.21 & 1.1474 \\
BN2 & 7.1 & 10.0 & 90.90 & 0.00 & 0.64 & 8.45 & 1.25 & 1.1573 \\
BN3 & 7.3 & 10.0 & 88.42 & 0.00 & 0.00 & 11.58 & 1.24 & 1.3180 \\
BN4 & 7.0 & 10.0 & 92.20 & 0.00 & 0.00 & 7.80 & 0.16 & 1.4435 \\
BS1 & 8.1 & 9.7 & 58.26 & 3.33 & 27.79 & 10.62 & 0.16 & 1.0284 \\
BS2 & 7.6 & 9.9 & 76.11 & 1.57 & 8.17 & 14.15 & 0.93 & 1.2923 \\
BS3 & 7.6 & 9.9 & 76.96 & 0.00 & 11.73 & 11.31 & 1.13 & 1.0580 \\
BS4 & 7.2 & 9.9 & 92.24 & 0.00 & 2.23 & 5.54 & 0.22 & 1.0427 \\
Nucleus & 8.1 & 9.9 & 53.50 & 14.83 & 15.33 & 16.33 & 1.12 & 1.4761 \\
\enddata
\tablecomments{Columns: \\
(1): Positions of spectrum used, which are marked in Figure~\ref{fig7}. \\
(2): the average age weighted by Luminosity \citep{Cid05}. \\
(3): the average age weighted by stellar mass \citep{Cid05}. \\
(4)-(7): the percentage of stellar population with different ages weighted by luminosity. The ages from (4) to (7) are Burst $<$ 10$^8$yr, Young 10$^8 - 10^9$yr, Median 10$^9 - 10^{10}$yr, Old $>$ 10$^{10}$yr, respectively. \\
(8): the extinction estimated by spectral synthesis. \\
(9): the minimum variance of each spectral synthesis.
}
\label{tabl3}
\end{deluxetable}

\begin{deluxetable}{cccccc}

\tabletypesize{\scriptsize}
\tablecaption{Potential Remnants of Minor Merger}
\tablewidth{0pt}
\tablehead{
\colhead{Name} & \colhead{RA} & \colhead{DEC} & \colhead{L$_{X}$$^a$} & \colhead{$\sum$ SFR} & \colhead{Log$_{10}$ F(24$\mu$m)/F(8$\mu$m)} \\
\colhead{label} & \colhead{J2000.0} & \colhead{J2000.0} & \colhead{erg s$^{-1}$} & \colhead{M$_{\bigodot}$ yr$^{-1}$ kpc$^{-2}$} & \colhead{}
}
\startdata
X1 & 23:04:57.62 & +12:20:28.70 & 5.35E+040 & 0.02 & -0.12\\
X2 & 23:04:51.54 & +12:18:26.40 & 1.52E+040 & 0.23 & -0.12\\
X3 & 23:04:52.10 & +12:19:42.40 & 5.69E+039 & 0.05 & -0.09\\
X4 & 23:04:56.82 & +12:19:28.80 & 1.83E+040 & ... & ...\\
H1 & 23:04:54.49 & +12:18:21.32 & ... & 0.36 & -0.08\\
H2 & 23:04:51.84 & +12:19:36.43 & ... & 0.51 & -0.04\\
H3 & 23:04:52.67 & +12:19:49.45 & ... & 0.33 & 0.11\\
H4 & 23:04:53.00 & +12:19:59.15 & ... & 0.04 & 0.04\\
H5 & 23:04:56.66 & +12:19:41.13 & ... & 0.03 & -0.09\\
L  & 23:04:56.66 & +12:19:41.13 & ... & 0.31 & 0.31\\
C$^b$  & 23:04:56.69 & +12:19:21.80 & 2.95E+040 & 0.89 & -0.77
\enddata
\tablecomments{\\
${}^a$ Including the emission from 0.2keV to 12keV. \\
${}^b$ The nucleus of NGC 7479. It is not the remnant, it is listed here just to be compared with the potential remnants.}
\label{tabl4}
\end{deluxetable}
\end{document}